\documentclass[twoside,twocolumn]{article}

\usepackage{blindtext} 
\usepackage{graphicx,graphics,epsfig} 
\usepackage[sc]{mathpazo} 
\usepackage[T1]{fontenc} 
\usepackage{microtype} 
\usepackage{cite}
\usepackage[english]{babel} 

\usepackage{float}
\usepackage[hmarginratio=1:1,top=32mm,columnsep=20pt,hmargin = 1.5cm  ]{geometry} 
\usepackage[hang, small,labelfont=bf,up,textfont=it,up]{caption} 
\usepackage{booktabs} 

\usepackage{lettrine} 

\usepackage{enumitem} 
\setlist[itemize]{noitemsep} 

\usepackage{abstract} 

\usepackage{titlesec} 
\titleformat{\section}[block]{\Large	\bfseries}{\thesection.}{1em}{} 
\titleformat{\subsection}[block]{\large	\bfseries}{\thesubsection.}{1em}{} 

\usepackage{fancyhdr} 
\usepackage{titling} 

\usepackage{hyperref} 
\usepackage[none]{hyphenat}

\pretitle{\begin{center}\Huge\bfseries} 
\posttitle{\end{center}} 
\title{PCrystalX - Web Application} 
\author{%
\textsc{Pablo L. Bernardo}\\
\normalsize State University of Northern Rio de Janeiro – UENF \\ 
\normalsize \href{mailto:pablolb@uenf.br}{pablolb@uenf.br}\\ 
\normalsize https://pcrystalx.uenf.br/ \\
}
\date{\today} 
\renewcommand{\maketitlehookd}{%
\begin{abstract}
\noindent PCrystalX - Web is a web application developed to be easy and fast to use. The purpose of the software is that researchers from any area are able to use two methods to analyze an X-ray diffraction pattern more straightforwardly. With the full width at half maximum (FWHM) as a function of 2$\theta$ for each peak, it is possible to estimate the crystallite size by the Scherrer equation and apply the Williamson-Hall methods. PCrystalX - Web is a simplified version of the PCrystalX Software \cite{pcrystalx} program, which is still under development.

\end{abstract}
}

\begin{document}

\maketitle


\section{Introduction}

X-ray diffraction is one of the oldest methods used by researchers for the structural characterization of various crystalline materials. However, the analysis of a diffractogram is not trivial. Although there are several programs for the refinement of data obtained by X-ray diffraction to describe the crystal structure of any compound, some programs require a high degree of crystallographic knowledge, complex interfaces to use, and some with expensive licenses. Most of these programs focus on the Le Bail and/or Rietveld refinement methods, where it is possible to refine the lattice parameters, constants U, V and W of the Caglioti equation to define FWHM as a function of the angle $\theta$, atomic positions, among many other parameters. Some examples are: FullProf Suite \cite{rodriguez2001fullprof}, Profex \cite{doebelin2015profex}, GSAS \cite{toby2001expgui}, TOPAS \cite{coelho2018topas}, HighScore Plus \cite{degen2014highscore}, and others.

PCrystalX - Web aims to be a program that will help researchers both in the analysis of raw data and in the analysis of data obtained through structural refinement. The main idea is to offer a program so that researchers from different areas can use it easily, and in some cases, without the need to carry out a structural refinement to obtain an interesting result from the X-ray data analysis. With PCristalX - Web it is possible, without the need for structural refinement, to estimate the crystallite size by the Scherrer equation and the crystallite size and strain by the Williamson-Hall plot. Furthermore, it is possible to use the parameters U, V and W, of the Caglioti equation, to apply the instrumental correction in the analysis (when a standard sample is measured).

\begin{figure}[ht]
\centering
\includegraphics[width=9cm, height=5cm ]{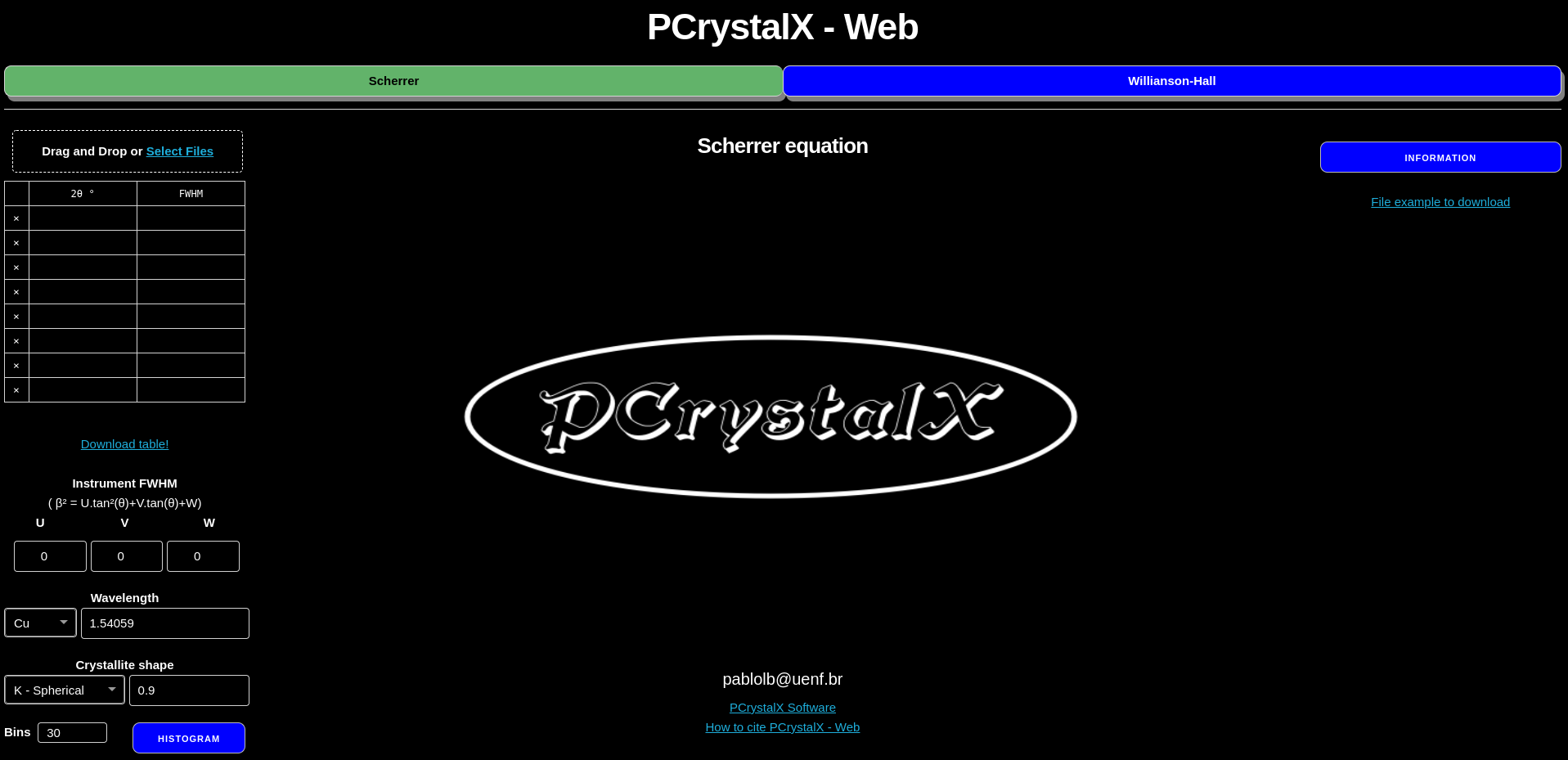} 
\caption{PCrystalX - Web page: https://pcrystalx.uenf.br/}
\label{fig1}
\end{figure}

Figure \ref{fig1} shows the first version of PCrystalX - Web application. Currently, it is possible to import data in txt and csv format or just type the values of $2\theta$ and FWHM into the table. The program was written using Python 3.9 and Dash framework \cite{dash}. The program is hosted on a server at the State University of Northern Rio de Janeiro – UENF, Brazil  (https://pcrystalx.uenf.br/) and is distributed online under the conditions of the GPL Open Source License (Version 3).

\clearpage

\section{General PCrystalX - Web functionality}

\subsection{Caglioti equation}

\begin{figure}[ht]
\centering
\includegraphics[width=8.5cm, height=6cm ]{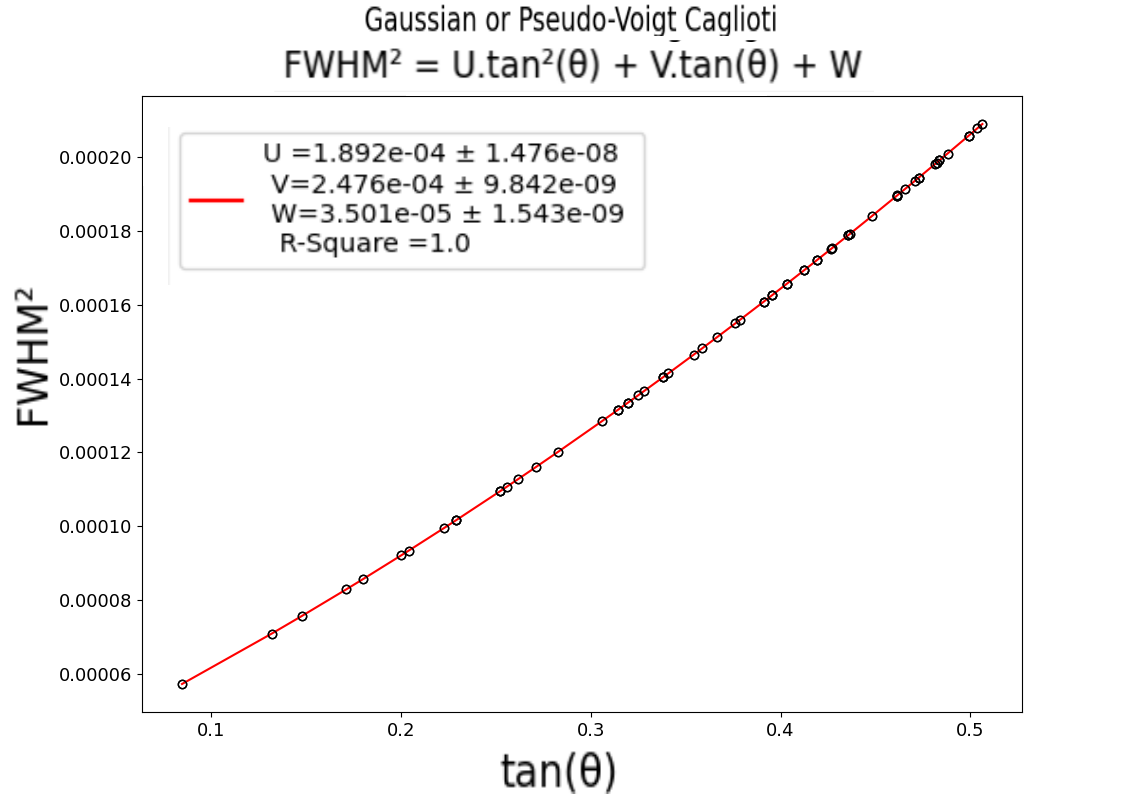} 
\caption{Parameters U, V and W obtained from the Caglioti equation in PCrystalX Software \cite{pcrystalx}}
\label{cagli}
\end{figure}

The observed diffraction peak profile is a convolution of the instrumental profile and the specimen profile. Once the user has the 2$\theta$ and FWHM ($\beta _i$) for each peak from standard specimen data (for example, $LaB_{6}$), it is possible to obtain U, V and W parameters from the Caglioti equation \cite{caglioti1958choice} for Gaussian and Pseudo-Voigt profiles (possible with PCrystalX Software, Figure \ref{cagli}).

When these parameters are obtained, we can use them to perform the instrumental correction using the Caglioti equations:

\vspace{0.2cm}

\begin{center}
\textbf{Gaussian  or  Pseudo-Voigt: }
\end{center}

\vspace{-0.6cm}

\begin{equation}
\beta _i^2  = \quad U.tan^2\theta  +  V.tan\theta  +  W
\end{equation}

\vspace{0.2cm}

PCrystalX - Web uses the integral-breadth ($\beta$)  method to perform the instrumental correction. For each diffraction peak, the integral-breadth for Gaussian or Pseudo-Voigt profiles is given by:

\begin{equation}
\beta^2 _s = \beta^2 _o - \beta^2 _i 
\end{equation}

 \vspace{0.3cm}

Where $\beta _s$, $\beta _o$ and $\beta _i$ are the integral-breadth of a specimen, observed and instrumental profiles. Initially, the values of U, V and W are assumed to be zero. The program will assume that $\beta _i$ = 0 and $\beta _s$ = $\beta _o$. If the user has the values of U, V and W, these can be used to perform the calculation. If any value is incompatible, a warning window will appear on the page as "\textit{U V W values not supported.}''

\subsection{Scherrer equation}

In X-ray powder diffraction, a family of planes produces a peak only at a specific angle 2$\theta$. Scherrer equation \cite{scherrer1918gottinger,patterson1939scherrer} can be used to estimate the crystallite size for each peak ($L_{hkl}$). The crystallite size calculated from the Scherrer equation is given by:

\begin{equation}
L \quad = \quad \frac{K_s.\lambda}{\beta_s.cos\theta}
\end{equation}

\vspace{0.2cm}

Where $K_s$ is a constant that depends on the shape of the crystallite, $\lambda$ is the wavelength, $\beta _s$ is the corrected line broadening at FWHM and $\theta$ is the scattering angle in radians. With 2$\theta$ and FWHM data, PCrystalX - Web can estimate the crystallite size for each peak and then obtain the average crystallite size ($L_m$). 

\begin{figure}[ht]
\centering
\includegraphics[width=9.5cm, height=6cm ]{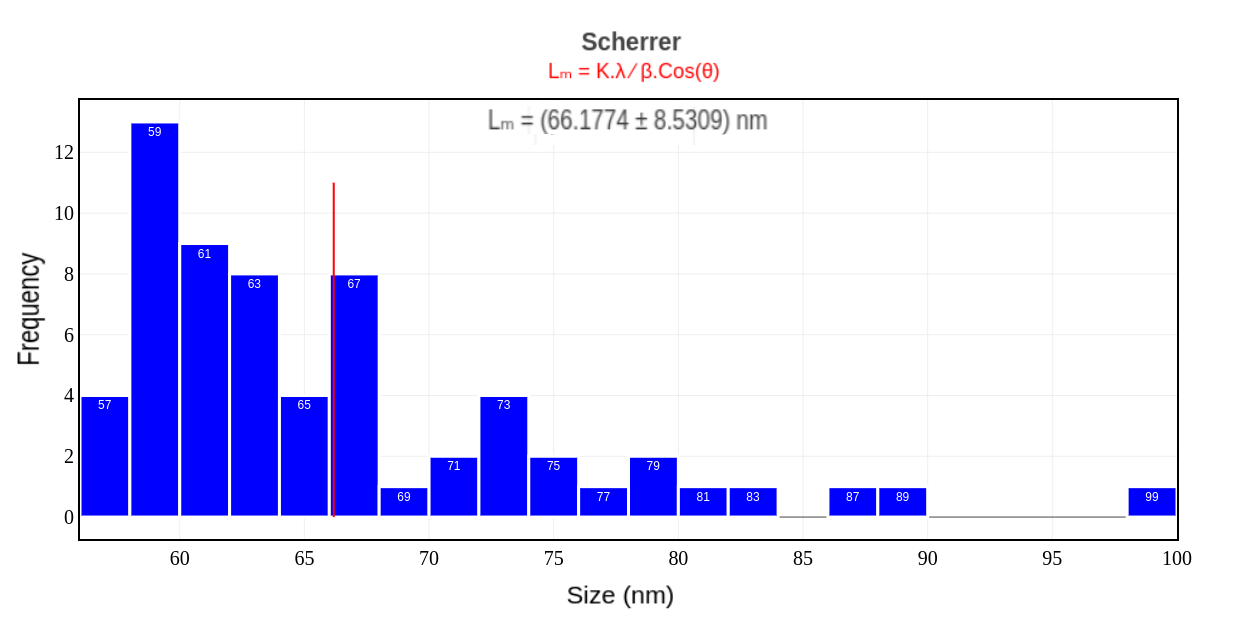} 
\caption{Histogram of crystallite size by  Scherrer equation}
\label{schee}
\end{figure}

The user can choose the wavelength and crystallite shape value, and it is possible to consider the parameters U, V and W to apply peak correction. A histogram plot will be generated with crystallite size values for each peak and the value of $L_m$ (Figure \ref{schee}).  If the value of the wavelength, crystallite shape, or bins is none, a warning window will appear on the page.

The user can change the histogram bins values and download the figure plot in png format. Furthermore, it is possible to export a result table containing the 2$\theta$, FWHM,  $L_{hkl}$  and $\delta L_{hkl}$ data in txt format.

\subsection{Williamson-Hall method}

In X-ray diffraction, a perfect crystalline sample produces sharp peaks. The absence of perfect crystallinity leads to a broadening of the diffraction peaks and we can consider two contributions to this broadening: crystallite size and lattice strain. Scherrer equation does not consider the lattice strain. However, G.K.Williamson and his student, W.H.Hall in 1953 \cite{williamson1953x} explained how crystallite size and strain can broaden the powder diffraction peak.

They assumed that the total broadening is due to the size broadening $\beta _L$ and the strain broadening $\beta _e$. So, $\beta _{Tot} = \beta _{L} + \beta _{e}$, and the Williamson-Hall equation is given by:

\vspace{-0.2cm}

\begin{equation}
\beta_{L} \quad = \quad \frac{K_s.\lambda}{Lcos\theta} \quad 
\end{equation}

\vspace{-0.2cm}

and

\vspace{-0.2cm}

\begin{equation}
 \beta_{e} \quad = \quad 4e.tan\theta
\end{equation}

\vspace{-0.2cm}

then,

\vspace{-0.3cm}

\begin{equation}
\beta_{Tot}.cos\theta \quad = \quad \frac{K_s.\lambda}{L} \quad + \quad 4e.sin\theta
\end{equation}

\vspace{0.2cm}

Where $e$ is the strain induced in powders. With 2$\theta$ and FWHM data for each peak, the user can choose the wavelength, crystallite shape value, and is possible to consider the parameters U, V and W to apply peak correction to estimate the crystallite size and strain. If the value of the wavelength, crystallite shape, or bins is none, a warning window will appear on the page.

\begin{figure}[H]
\centering
\includegraphics[width=9.5cm, height=6cm]{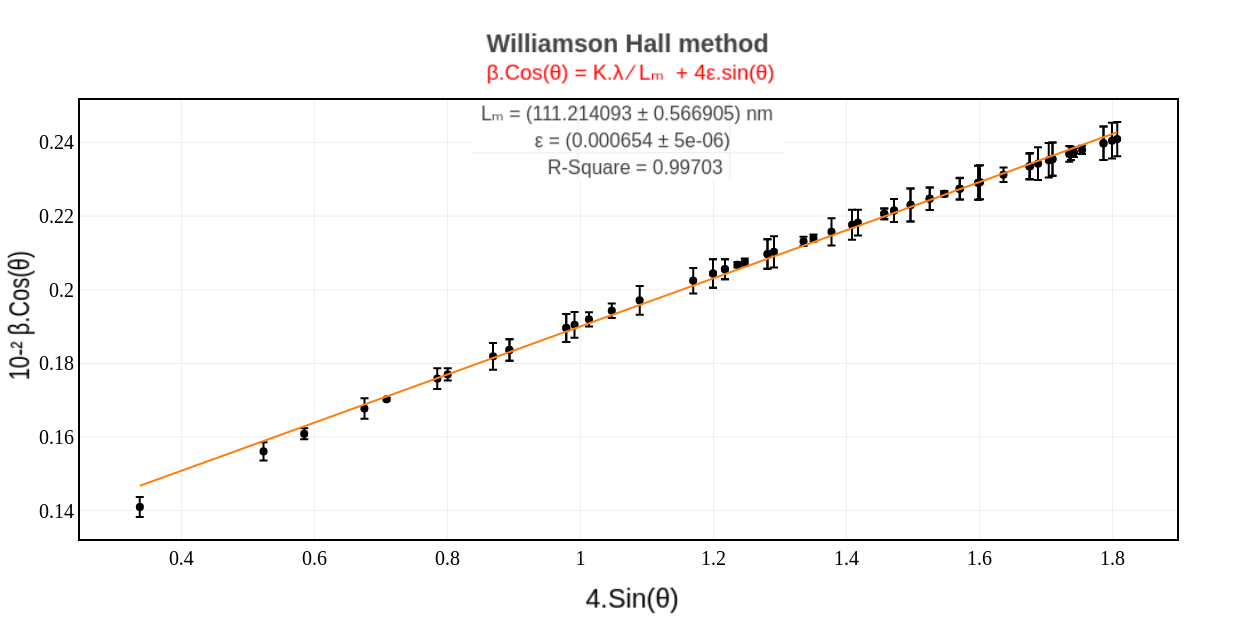} 
\caption{Williamson - Hall analysis: determination of crystallite size and strain}
\label{wil}
\end{figure}

Thereby, a $\beta_{s}.cos\theta$ vs $4.sin\theta$ plot will be generated and a linear fit will be performed automatically on the data. With this, we can obtain the average value of the crystallite size and the lattice strain value. The user can download the figure plot in png format (Figure \ref{wil}) and export a result table containing the $4.sin\theta$, $\beta_{s}.cos\theta$,  $\delta \beta_{s}.cos\theta$  and Fit data in txt format.

\section{Conclusions}

PCrystalX - Web is a web application developed to help researchers from different areas in the analysis of data obtained by X-ray diffraction, in particular, in the estimation of crystallite size and strain. This web application stands out as it is not necessary to download or install any program on a computer. The user just needs to access the internet through the computer, tablet, or smartphone to perform the calculations using the Scherrer equation and the Williamson-Hall method. In the future, new applications may be added.

\section{Acknowledgements}

We would like to thank all users who are helping to improve the PCrystalX-Web application by providing their suggestions. Also, a special thanks to Nikola Bernardo, Cintia Baltar, André Pecini, Wellington Wallace and André Rangel.

\bibliographystyle{unsrt}
\bibliography{ms}

\end{document}